\documentstyle[preprint,aps,psfig]{revtex}
\begin{document}
\title{Effects of thermal fluctuation and the receptor-receptor interaction
in bacterial chemotactic signalling and adaptation}
\author{Yu Shi
\footnote{
Email: ys219@phy.cam.ac.uk}}
\address{ 
Cavendish Laboratory, University of Cambridge,
Cambridge CB3 0HE, United Kingdom}
\maketitle
\begin{abstract}
Bacterial chemotaxis is controlled by the conformational changes of the 
receptors, in 
response to the  change of the  ambient chemical concentration.
In a statistical mechanical approach, the signalling due to the 
conformational changes is
a thermodynamic average  quantity, dependent  on  the temperature and the 
total energy of the system, including both  ligand-receptor 
interaction and receptor-receptor interaction.
This physical theory suggests to biology  a new understanding
of  cooperation in ligand binding and receptor signalling problems. 
How much experimental 
support of this approach can be obtained  from  the currently available data?
What are the parameter values? What is the  practical information
for experiments? 
Here we make comparisons between the theory and recent experimental 
results.  Although currently   comparisons can only be 
semi-quantitative or qualitative,  consistency is clearly shown.
The theory also helps to sort a variety of data.
\end{abstract}


\section{Introduction}

Bacterial chemotaxis refers to the phenomenon that a bacterium such as 
{\it Escherichia coli} swims towards higher concentration of
attractant or  lower concentration of
 repellent~\cite{berg1993,stock,falke,berg}.
With the switching rate determined by the   change of the 
ambient chemical concentration, the motors of
the bacterium  switch between counterclockwise and
clockwise rotations, consequently the bacterium   switches between 
tumbling and running. 
The ratio between the frequencies of the two rotation modes is determined
by the rate at which  kinase CheA phosphorylates CheY, which binds the 
base of a  motor. CheA phosphorylation rate  is regulated by   
the receptor  conformational state, which is   influenced by ligand binding. 
The receptors are dimeric and   joined to 
a CheA dimer by a CheW dimer, furnishing a signalling complex.
Hence  a receptor dimer can be regarded  as a basic unit, as supported by
the finding that a receptor  dimer with a damaged subunit can still 
work~\cite{gardina}.  
Because of thermal fluctuation, even in the absence of 
ligand binding, or in a fully adapted situation,
there is still a certain probability distribution for a  receptor 
dimer to be in different  conformational states; 
microscopically a receptor dimer
stochastically flips between the two states. 
Attractant binding changes the probability distribution, causing the
receptor dimer to be more likely in the 
state corresponding  to the lower  CheA   phosphorylation rate.
On a longer time scale, after an initial response to ligand concentration
change, the activity of the system returns to the pre-stimulus level. 
A careful consideration of such a basic  picture already  finds the   
ideas of statistical mechanics   necessary: with the presence of thermal 
fluctuation, it is the probability  distribution of   the
conformational states of the receptors
that is monitored by ligand concentration change and determines the
motor rotation bias.  However, it  seems that this point is not  
universally appreciated in biological literature. 

The chemotactic response is very sensitive~\cite{segall}. It 
 had been  conjectured 
that there might be cooperation between receptors or the signalling
complexes  so that the signal could be 
amplified~\cite{borkovich,falke}.
The fact that most of the receptors cluster 
together at a pole of the cell  provides further  clues for cooperation  
between  receptors~\cite{parkinson,bray}.  More importantly,
it was found experimentally  that
the clustering of receptors  is  not   favorable for counting
statistics and  that the receptor cluster does not favor a special end
of the cell~\cite{berg2}. This is an indication 
that there is  a special reason for clustering, 
which may well be to make the  receptor-receptor interaction possible.

With  a  detailed analysis 
on the experimental findings, I suggested the
    possible existence of interaction between neighboring
    receptor dimers and constructed
a statistical mechanical theory  to provide a picture 
of how the receptors cooperate through physical interaction and how
the thermal fluctuation makes statistical mechanics important in the 
signalling process~\cite{shi1,shi2}.
In our model, we combine cooperativity and feedback to 
   account for the sensitivity and adaptation.
As will be stressed   here, the first message  
from  our   approach  is an emphasis on thermal fluctuation. 
In a cell, the energy scale is comparable with the thermal fluctuation.
Moreover, thermal fluctuation  helps  to distinguish different stimuli.
Because of
large separation of time scales, the  thermal fluctuation   can be treated
as quasi-equilibrium, so equilibrium  statistical mechanical can   give a 
reasonable response-stimulus relation. Hence  the basic elements
of our  theory 
is useful  no matter whether there is interaction  between receptor dimers. 
The second message of our  theory  is that the anticipated
cooperation is  just physical receptor-receptor 
interaction between  neighboring receptor dimers.
Therefore the conformational state  of a 
receptor dimer is not only influenced by  ligand binding of itself, but also
by the receptor-receptor interaction which  depends
on the  conformations of the two neighboring receptor dimers. 
The third message is that the large separation of time scales leads  to
a complementary usage of equilibrium statistical mechanics  in calculating 
the  response on a shorter time scale and
a non-equilibrium description of the adaptation 
on a longer time scale. Dynamics on the longer time scale
determines  whether  randomness of 
ligand binding  is quenched or annealed  on the shorter  time
scale of quasi-equilibrium state, as will be elaborated later on.  
In the high temperature limit, this does not make a difference on the average
signalling.  

Recently there appeared some  experimental data which are more directly 
relevant for the many-body nature of the receptor cluster and 
the possible cooperation \cite{jasuja1999b,bornhorst,sourjik}.
Therefore it is interesting and important
to make comparisons between the theory and the
experimental results, testing  the theory on  one hand,
and providing some  information on what experiments are wanted on the 
other hand.  However,
we do not expect the model in  current form  to   perfectly fit 
everything  on this complex system. 
Rather, what we provide is a theoretical framework, on which 
refinements are possible. For example,  we have only considered  
the cooperation between the receptor dimers, while extensions  to 
possible cooperation  among other  components at later stages of the
signalling process, for example, CheA, CheY, CheZ and
the switch complex, is  straightforward if sufficient  information
is available.
The idea of receptor-receptor interaction broadens the view on cooperation,
which previously  largely meant  the existence of more than
 one binding sites, as described by
the model presented by Hill a century ago~\cite{hill}.
For simplicity, we try to preserve the scenario of  one binding site, 
while  extension  to the situation with  more binding sites
is straightforward if necessary.
Our strategy is to start with the minimum model, which yet explains the most 
essential features. 

With   improvement and  simplification,
we  first describe the  theory. Then we make  comparisons with the 
experimental results, followed by summary and discussions. 

\section{ Theory}

Consider a lattice of receptor dimers, as shown in Fig. 1. 
Let the coordinate number be $\nu$, which is $6$ for a honeycomb lattice
and is $4$ for a square lattice. The exact coordinate number
in reality is subject to experimental investigations. 
The behavior of the system is determined by its energy function,
or Hamiltonian, which can be written as
\begin{equation}
{\cal H}(t)\,=\,-\sum_{<ij>}T_{ij}V_{i}V_{j}-\sum_{i}H_{i}V_{i}
+\sum_{i}W_{i}V_{i}.\label{hamiltonian}
\end{equation}
$V_i$ is a variable characterizing the conformation of receptor dimer
$i$. It may be interpreted as the position of
 the receptor molecule with respect
to a certain equilibrium position.
In the popular  two-state approach, $V_i$
assumes one of two values
$V^0$ or $V^1$.  $H_i$ is the influence, or force, 
 due to  ligand binding and the modulation of methylation level.
$H_i=0$  if there is 
no ligand binding, while $H_i=H$ if there is a ligand binding.
$-H_iV_i$ is the energy due to ligand binding, hence
 ligand binding causes the energy difference 
between the two conformations to make a shift of  $H(V^1-V^0)$. 
$W_i (V^0-V^1)$
is the original  energy difference between the two conformations. 
$\langle ij\rangle$
denotes nearest neighbouring pairs,    
$-T_{ij}V_{i}V_{j}$ is the interaction energy  between
the neighboring receptor dimers. 

For convenience, defining 
$S_{i}\,=\,2(V_{i}-V^{0})/\Delta V-1$, with $\Delta V=V^{1}-V^{0}$,
one transforms  the Hamiltonian  to  
\begin{equation}
{\cal H}(t)\,=\,-\sum_{\langle ij\rangle} J_{ij} S_{i}S_{j}-
\sum_{i}B_{i}(t)S_{i}+\sum_{i}U_i S_{i},
\label{ising1}
\end{equation}
where $S_i=1,-1$ represents the two conformational states of
the receptor dimer at site $i$,  
 $J_{ij}\,=\,T_{ij}\Delta V^2/4$, $B_{i}\,=H_{i}\Delta V/2$,
$U_i=\Delta V W_i/2-\Delta V^2 \sum_jT_{ij}$. 
We refer to  $B_i$ as  ``field''.
For simplicity, it is  assumed that
$J_{ij}=J$  and $U_i=U$  are  independent of   $i$ and $j$.
$B_i=0$ if there is 
no ligand binding, while  $B_i=B=H\Delta V/2$ if there is a ligand binding.
Hence energy difference due to ligand binding 
between the two conformations are $2B$. 
$US_i$  represents the original energy 
in the absence of ligand binding. 
Eq. (\ref{hamiltonian}) and (\ref{ising1}) can be justified as follows.
It is reasonable to assume an interaction  energy  proportional 
to $(V_i-V_j)^2$, which  can  be reduced to $-T_{ij}V_{i}V_{j}$,  with
constant terms neglected and 
the terms proportional to $S_i$ or $S_j$ included in $\sum_i U_iS_i$.  
This assumption is simple enough to allow a feasible treatment which 
yet captures the essential features.  

From now on, we   focus on Eq. (\ref{ising1}). 
Suppose that before  time $t=0$, there is no  ligand  binding   the 
system, or the system is fully adapted though it is bound to ligands.
Hence  $B_i(t<0)=0$.
Afterwards,  at   time $t=0$,  the occupancy, i.e. the fraction of receptor
dimers with ligands bound, changes to $c$. Hence the occupancy change is
$\delta c=c$.  This means $B_i(t=0)=B_i^0$, with
\begin{equation}
B_i^0=\left\{\begin{array}{l}
B, {\rm with \mbox{ } probability} \mbox{ } c\\
0, {\rm with \mbox{ } probability} \mbox{ } 1-c
\end{array}\right.
\end{equation}
The occupancy $c$ is determined by the ligand  concentration $L$,
$c=L/(L+K_d)$, where the dissociation constant 
 $K_d$   is   on a time scale during which the
receptor has undergone many flips between different conformations, hence
it is an  average and phenomenological quantity. 

On the other hand, through the modulation of methylation level
by CheB and CheR, there is a negative feedback from the receptor
state $S_i$ to the field $B_i$, with a time delay $t_r$. A simple
quantitative representation  of this feedback is
\begin{equation}
\frac{dB_{i}(t)}{dt}\,=\,-\sigma [S_{i}(t-t_r)-m_0], \label{cb1}
\end{equation}
where  $\sigma>0$, $m_0$ is the pre-stimulus average of $S_i$.
Precise forms of both
the energy function and the feedback are, of course, subject to experimental 
investigations.   It seems  that in biological world, 
feedback is a ubiquitous way to
achieve adaptation and preserve sensitivity of response. 

A remarkable feature of this system is the large separation of time scales.
Ligand binding and conformational change occur  within
only   millisecond,  while overall
time needed to complete the adaptation, through
the  slow modulation of methylation level,
is on  the time scale of  many seconds to  minutes~\cite{bieman,stock}. 
We note that in most cases,  ligand debinding is on a much longer 
time scale than ligand  binding, seen as follows. 
Consider the kinetics of the following reaction
\begin{equation}
L+R 
\rightleftharpoons  R_L,
\end{equation}
where $R$ represents the receptor  without ligand binding, while $R_L$ 
 represents the liganded receptor. $k_+$ and $k_-$ are reaction
rates  for  the binding and debinding, respectively. The ratio between
the time scales of debinding and binding is 
$k_{+}L/k_{-}\equiv L/K_d$, where $K_d$ is the dissociation constant. A typical
value is 
 $K_d \sim 1.2\mu M$~\cite{stock}.
 Usually,  $L$ is   much larger, so
the debinding time scale is much longer than the time scale of ligand binding
and receptor conformational change. In extreme cases,  when $L$ is 
comparable with  $K_d$, debinding time scale is comparable to binding
time scale.  

With the large separation of time scales, the treatment within  the above 
framework  becomes  easier. One may 
discretize the time on the scale of adaptation, according to the feedback
delay time. $t$ is thus replaced by an integer $\tau$, which is 
the integer part of $t/t_r$. On the other hand,{\em 
 each instant $\tau$ is still
very long compared with the time scale of conformational change}. 
Hence the activity at each $\tau$ is an average quantity $m(\tau)$, which 
 can be calculated from the Hamiltonian in  (\ref{ising1})
 by standard methods of equilibrium statistical mechanics.
The average activity $m$  is just  on the time scale of the
measurement of the macroscopic  quantities such as motor bias, 
longer  than the very short period in which the receptor is in either
of the two conformations, but shorter than the adaptation time.
In making the average,
an important thing is that the randomness of the
field is usually  quenched since $L>>K_d$, and is annealed otherwise.
In fact we obtain a generalized version of the so-called
random-field Ising model; in the   conventional
random-field Ising model,
the average field vanishes,
but it is generically non-zero in our model. On the long time scale, 
the field changes  because of feedback. It can be expressed as
$B_i(\tau)=B_i^0+M(\tau)$, where
 $M(\tau)$ is an induced field  due to methylation modulation,
\begin{equation}
M(\tau)\,=\, -\sigma \sum^{\tau -1}_{k = 0} [m(k)-m_0] \label{mtau}.
\end{equation}
Before stimulation, $m(\tau<0)=m_0$ is determined by 
$U$.  If and only if  $U=0$, 
$m_0=0$,  which means that each  receptor dimer is
in either  of the two conformational states  with equal probability.

In most cases, the  randomness of $B_i^0$ is quenched,
the general relation between $m(\tau)$ and  $\delta c$ is then
\begin{eqnarray}
m(\tau)=&
\frac{2 \delta c}{1+\exp[-2\beta (\nu Jm(\tau)-\theta(\tau-1)
\sigma\sum_{k=\tau_0}^{\tau-1}(m(k)-m_0)+ U+B)]} \nonumber \\
&+\frac{2(1-\delta c)}{1+\exp [-2\beta(\nu Jm(\tau)-\theta(\tau-1)
\sigma\sum_{k=\tau_0}^{\tau-1}(m(k)-m_0)+U)]}-1,
\label{mag}
\end{eqnarray}
where $\beta=1/k_BT$, 
the step function  $\theta(x)$ is $1$ if $x \geq 0$,  and is $0$ otherwise.
On the other hand,
when  ligand concentration is comparable with  $K_d$,
the randomness of $B_i^0$ is annealed. Then  it can be found that
\begin{equation}
m=\frac{\delta c [e^{\beta(f(m)+B)}-e^{-\beta(f(m)+B)}]
+(1-\delta c)[e^{\beta f(m)}-e^{-\beta f(m)}]
}{
\delta c[e^{\beta(f(m)+B)}+e^{-\beta(f(m)+B)}]
+(1-\delta c)[e^{\beta f(m)}+e^{-\beta f(m)}]
}, \label{ann}
\end{equation}
where 
$f(m)=\nu J m- \theta (\tau -1) \sigma \sum_{k=0}^{\tau -1} m(k)+U$.

$m(\tau)$ vs. $c$ relation  corresponds to  the   response-stimulus relation.
After the step increase  at $\tau=0$,
 $m(\tau)$ always decreases   towards 
the pre-stimulus value $m_0$. This explains  the robustness of 
 exact adaptation~\cite{alon}.
 In practice 
the adaptation time is obtained when $m-m_0$ reaches  the detection
threshold $m^*$.

The results  can be simplified under 
 the condition that the thermal fluctuation  is so strong that 
 $\beta\nu J$ and $\beta B$ are not  large.  
Then both Eq. (\ref{mag}) and Eq. (\ref{ann}) 
   can be simplified to
\begin{equation}
m(\tau\geq 0) -m_0 = \frac{\beta B\delta c}{1-\beta\nu J}
\left(1-\frac{\beta\sigma}{1-\beta\nu J}\right)^{\tau},
\label{simp}
\end{equation}
with 
\begin{equation}
m_0 = \frac{\beta U}{1-\beta\nu J}.
\label{m0}
\end{equation}
$1-\beta\nu J$ represents the enhancement of response compared
with non-interacting scenario. 

One may obtain the adaptation time $t^*$, after which 
$m-m_0$ is less than   the detection threshold $m^*$:
\begin{equation}
\tau^*=\frac{\log \delta c+ \log(\frac{\beta B}{1-\beta\nu J})-\log m^*}
{-\ln(1-\frac{\beta \sigma}{1-\beta\nu J})}.
\label{time}
\end{equation}
$m^*$ can be related to the lower bound of detectable occupancy change,
$\delta c^*$,  by
\begin{equation}
m^* = \frac{\beta B\delta c^*}{1-\beta\nu J},
\label{m*}
\end{equation}
hence 
\begin{equation}
\tau^*=\frac{\log\delta c- \log\delta c ^*}
{-\ln(1-\frac{\beta \sigma}{1-\beta\nu J})}.
\label{time2}
\end{equation}

At exact adaptation, 
setting $m(\tau)=m_0$, one may obtain the total
induced field due to methylation modulation as $M^*=Bc$. Then for
 the next stimulus, suppose  that the 
occupancy changes from $\delta c$ 
  to $\delta c+\Delta c$ at a later time $\tau_1$, 
 it can be found that
the result  with the occupancy  $\delta c+\Delta c$ 
 and the induced field 
$M^*$ is the same as that for the situation in which
the occupancy is $\Delta c$ and there is no induced field.  That is to say, 
the previous occupancy change has been canceled by the induced filed  $M^*$, 
therefore the fully adaptation with ligand binding
 is equivalent to no ligand binding.  
So $m(\tau\geq \tau_1)$ is given by the above relevant equations
with 
$\tau$ changed to $\tau-\tau_1$, and $\delta c$ substituted by $\Delta c$.
 One can thus simply forget 
the pre-adaptation  history, and re-start the application of 
the above formulation with  $\tau_1$ shifted to $0$.   
The cancellation holds exactly only  under the assumption 
of small $\beta\nu J$ and $\beta B$, which is likely the reality.
The finiteness of detection threshold further 
widens the practical  range of its  validity. 

\section{ Comparisons between experiments and the theory}

\subsection{ Clustering.}
 
The clustering has recently been experimentally  studied in greater
 detail~\cite{sourjik}.
 The observed  clustering of receptors and the
co-localization of the CheA, CheY, and CheZ with the receptors
 is  a favor for the effects of interactions. 
An {\em in vitro}
receptor lattice formation was also observed~\cite{ying}.

\subsection{ Response-stimulus relation.}

A basic prediction of our theory is the response-stimulus relation. 
Note that the time scale of the response, corresponding to $m$ 
in our theory, is longer than 
the very short lifetime of a specific conformation, but
 is only transient on the time scale of the adaptation process.  
An interesting  thing is that $m$ in our theory  is  measurable.
Motor rotation bias was 
measured~\cite{jasuja1999b}. From this result we can obtain $m$, as follows.
 The motor bias is 
\begin{equation}
b=f_{ccw}/(f_{ccw}+f_{cw}),
\end{equation}
 where $f_{ccw}$ and 
$f_{cw}$ are rates of counterclockwise and clockwise rotations, respectively.
Suppose the value of $b$ is $r_{1}$ for  conformational 
state $1$, and is $r_{-1}$ for conformational state $-1$. 
Then the average bias is:
\begin{equation}
\overline{b} = r_{1} x+r_{-1}(1-x),
\end{equation}
where $x$ is the average fraction 
 of receptors with state $1$. $x$ is related to $m$ by
 $m=x-(1-x)=2x-1$.
So if we know $r_{1}$ and $r_{-1}$, we can obtain  $m$ from 
$\overline{b}$.   However, there seems to be  no investigation on 
$r_1$ and $r_{-1}$. A simple assumption which is often  implicitly 
 assumed   is that  
  $r_{1}=1$, $r_{-1}=0$, 
i.e.  state $1$ corresponds
to counterclockwise rotation,
while 
 state $-1$ corresponds to clockwise rotation. We follow this assumption here. 
But it should be kept in mind that an experimental 
investigation on  $r_{1}$ and $r_{-1}$ would be very valuable. 
Therefore, for the time being, we use
\begin{equation}
\overline{b}   =\frac{m+1}{2}, \label{mb}
\end{equation}
Thus from the pre-stimulus value of $\overline{b}$,
 one may determine $m_0$, and thus
$\beta U$. An empirical formula is $\overline{b}=1-0.0012(rcd-360)$, where
$rcd$ is the absolute angular rate of change of direction of the
cell centroid in $degree\cdot s^{-1}$~\cite{jasuja1999b,khan1993}.
 From~\cite{khan1993},
 the pre-stimulus value of $rcd$ is known as  $\sim 600$,
so the pre-stimulus value of $\overline{b}$ is $\sim 0.712$. Hence
\begin{equation} 
m_0 = \frac{\beta U}{1-\beta \nu J}\approx 0.424. 
\label{m0v}
\end{equation}

The occupancy change used in~\cite{jasuja1999b}
was calculated from the ligand  concentration 
under the assumption  that  the ligand  randomly binds   one of 
two possible binding sites: 
in addition to the site with  $K_d\sim 1.2 \mu M$,
as  widely acknowledged~\cite{bieman},
there is another site with  $K_d\sim 70 \mu M$.
This was based on an earlier  attempt to have a 
 better fitting for  the adaptation time~\cite{jasuja1999a}.
However, as said  above, we try  to make things as simple as possible
in the first instance, so prefer to  preserve the 
scenario of one binding site with  $K_d\sim 1.2 \mu M$.  
Actually with one binding site, as discussed later on, it seems that 
our theory  can fit the adaptation time by choosing appropriate
parameter values, thus  improve the coherence among  various data.
So  we should 
first transform  the values of  the occupancy change in~\cite{jasuja1999b}
 to the values  
one would have obtained  without the assumption of two binding sites.
 One has 
\begin{equation}
c_J=\frac{1}{2}(c_1+c_2),
\end{equation}
where $c_J$ represents the occupancy used by
Jasuja {\it et al.},
$c_1$ corresponds to dissociation constant $K_1=1.2 \mu M$,
$c_2$ corresponds to dissociation constant $K_2=70 \mu M$. 
From $c_l=L/(L+K_l)$, $l=1,2$, one obtains the change of the occupancy
\begin{equation}
\delta c_l=\frac{K_l\delta L}{(L+\delta L+K_l)(L+K_l)},
\label{dc}
\end{equation}
where $\delta L$ is the change of ligand concentration. 
Since $\delta L<<L$, one may obtain
$\delta c_1=2\delta c_J/(1+\alpha)$,
 where $\alpha \approx K_1(L+K_1)^2/K_2(L+K_2)^2$.
With $L\approx 10\mu M$, $\alpha\approx 1$, one has
 $\delta c_1\approx \delta c_J$.
 Therefore under this condition, 
we may simply  use the occupancy change  in~\cite{jasuja1999b}.
Eq. (\ref{mb}) leads to the relation between the initial 
change of $m$ and that of the motor bias, $\delta b$, 
\begin{equation}
\delta m=2 \delta \overline{b},
\end{equation}
where 
$\delta m= m(\delta c,\tau=0)-m_0$.

So the data in Fig. 3 of~\cite{jasuja1999b}  can be  transformed to 
$\delta m$ vs. $\delta c $ relation as shown in our Fig. 2.
Unfortunately,  it is  notable that 
the data is limited to very low values  of
occupancy change! Nevertheless, a qualitative 
fitting can be made. 
According  Eq. (\ref{simp}), setting  $\tau=0$,
we fit the data  with a straight line
$\delta m=a\delta c$, where
\begin{equation}
a=\frac{\beta B}{1-\beta\nu J} \approx 10.49
\label{bb}
\end{equation}
is  the slope of the fitting line.

\subsection{ Adaptation time.}

Eq.~(\ref{dc}) tells us that with a same concentration change, the 
occupancy change and thus 
the response {\em decreases} with the increase of  pre-stimulus 
ligand concentration. This is verified  by Fig. 7  of 
\cite{jasuja1999a}.
Eq.~(\ref{time}) predicts that the adaptation time
increases linearly with, but not proportional to,   the logarithm of
occupancy change. 
It had been thought that the adaptation time is proportional to the
occupancy change~\cite{berg3,spudich,jasuja1999a}. 
We found that a logarithmic relation is also consistent with the currently 
available data. As an example,  we examine the better set of the
data, the left plot (D-ribose),  in Fig.~4 of \cite{spudich}.
For accuracy, the two 
data points at the highest and lowest concentration changes
are dropped. This is  because they are at the detection limits, and  
they have  no recognizable  differences  in adaptation time   with the
 data points closest to them respectively,
although the values of  concentration  change are
quite different.  Moreover,
the adaptation time is recorded to be zero for
the two smallest values of concentration change, so 
the data point with the smallest concentration 
change should be ignored.  
Using $K_d=3\times 10^{-7}$ (no unit was given, but should be the same as
that of the concentration, so there is no problem in using it),
we transform the concentration to the occupancy.
The transformed data is shown in our  Fig.~3, with Fig.~3a the normal-normal
plot and Fig.~3b the normal-log plot. 
While  there could be a 
fitting with a proportional relation, as usually assumed, it is  at least 
reasonable to fit them with  a logarithmic relation, 
$t^*/0.354s\equiv\tau^*\cdot t_r/0.354s= (g\log_{10} \delta c+h)/0.354s$, with
$g= 95.151\times 0.354s=33.7s$ and $h=124.0574\times 0.354s=43.9s$. The factor
$0.354s$ comes from the data  normalization  in  \cite{spudich}, which is 
the percentage of  one of 
three maximum recovery times, $0.56m$, $0.58m$ and $0.62m$,
i.e. $35.4s$ on average.  
From Eq.~(\ref{time}), we have 
\begin{equation}
\frac{t_r}{-\log_{10} (1-\frac{\beta \sigma}{1-\beta\nu J})}=g. \label{t1}
\end{equation}
and
\begin{equation}
\frac{t_r[ \log_{10}\delta c^*]}
{\log_{10}(1-\frac{\beta \sigma}{1-\beta\nu J})}
=h. \label{t2}
\end{equation}
Using  $\delta c^*\approx 0.004$~\cite{jasuja1999a},
and assuming  $t_r\approx 0.1 s$, one finds 
\begin{equation}
\frac{\beta \sigma}{1-\beta\nu J}\approx 0.0068 \mbox{ }{\rm to}\mbox{ } 
0.013.
\label{bsigma1}
\end{equation}
where the first value is  estimated by using  (\ref{t1}), and the
second  by using (\ref{t2}). They are  close to each other,
 as  an indication of  the consistency of the theory.

Furthermore, 
our predicted  logarithmic relation may  explain the discrepancy 
in the analysis of the data in  Fig. 4 of~\cite{jasuja1999a}
about a   relation between the adaptation time and the concentration. 
The logarithm can simply  decrease
the predicted value of  adaptation time, without resorting to the assumption 
of  the existence of  two binding sites.  
We have tried to make a quantitative fitting for the data in
 Fig. 4 of~\cite{jasuja1999a}.  Using $K_d =1.2 \mu M$,
 we transform the ligand concentration to
the occupancy, as shown in  our Fig.~4.
To make  better use of the  data, we ignore data point for
$\delta c>0.95$, because the finiteness of detection threshold
may cause uncertainty in deciding the adaptation time; the data for 
$\delta c>0.95$ show too large variation for so close values of $\delta c$.
The fitting straight line is
$t^*\equiv\tau^*\cdot t_r= g\log_{10} \delta c+h$, with
$g= 156.3513$ and $h=114.9912$.
Using (\ref{t1}) and (\ref{t2}), and same
values of $\delta c^*$ and  $t_r$ as above,  one  finds 
\begin{equation}
\frac{\beta \sigma}{1-\beta\nu J}\approx 0.0015 \mbox{ }{\rm to}\mbox{ } 
0.0047.
\label{bsigma}
\end{equation}
Again, these two numbers are  close to each other. Moreover,
 (\ref{bsigma1}) and (\ref{bsigma})
are of the same order of magnitude,
 though they are obtained from  different sets of data. Even closer numbers
may be obtained by tuning the vale of $t_r$.

\subsection{  CheA activity.}

Another interesting and important experimental result  is on  the
relative  CheA activity, which has been 
analysed by using Hill model with a non-integer 
coefficient~\cite{bornhorst}. 
Here we examine  the data from the perspective  of our theory. 

Suppose $S=1,-1$ correspond respectively to  CheA activity
$A_1$ and $A_{-1}$. Then the average CheA activity is 
$\frac{1}{2}(A_1+A_{-1})+\frac{m}{2}(A_1-A_{-1})$. Consequently 
the relative CheA activity, as measured in \cite{bornhorst}, 
 is
\begin{equation}
R=\frac{(A_1+A_{-1})+(A_1-A_{-1})m(\delta c)}
{(A_1+A_{-1})+(A_1-A_{-1})m(\delta c=0)}
=1-F\frac{L}{L+K_d},
\end{equation}
where $F= \frac{a }{E+a\frac{U}{B}}$, with 
 $E=(A_{-1}+A{1}/(A_{-1}-A{1})>0$. Note that 
$A_{-1}>A_{1}$.  It is constrained that for attractant binding, 
$F\leq 1$,  since $R\geq 0$. 
Setting $F=0.95$ and $K_d=20\mu M$,
we obtain  a reasonable fitting to Fig. 1 of \cite{bornhorst},
as shown in our Fig.~4. Therefore
\begin{equation}
E\approx a(\frac{1}{0.95}-\frac{U}{B}), 
\label{fe}
\end{equation}
which, combined with Eqs.~(\ref{m0v}) and ~(\ref{bb}), implies  that  
the ratio between the two levels of CheA activity is
$A_{-1}/A_{1}\approx 164.77$. {\em
Very interestingly, this result of deduction 
 is in good  consistency with the 
available experimental information that this ratio is more than 
100}~\cite{stock}. Again, this is an indication of the
consistency of the theory.

We note that  there is discrepancy in the fitting. This may be
 because  of some other 
factors not considered here, especially,
 may be because the
correspondence between the receptor conformational state and 
CheA activity is  more complicated, in connection with $r_1$ and
$r_{-1}$ discussed above.   

\section{ Summary and discussions}

We suggest that statistical mechanics is useful and important in understanding
receptor signalling and  adaptation.  
We have made semi-quantitative  comparisons between the theory and
recent experiments to obtain estimations of parameter values.
  The 
thermal fluctuation in a cell is very strong, $k_BT\approx 4 pN\cdot nm\approx
0.025eV$, comparable  to the energy scale. So we 
simplify the formulation by using high temperature approximation. 
Then Eqs. (\ref{simp}) and (\ref{m0}) essentially 
contain all the information we need. $1-\beta\nu J$ characterizes  the 
enhancement of signalling by receptor-receptor interaction. 
With this  simplified  formulation, we look at
recent experimental results. Unlike a clean system usually studied in physics,
for  such a complex system, we do not expect the fitting to be
quantitatively  perfect. From the data on  pre-stimulus motor rotation 
bias~\cite{khan1993}, we obtain the pre-stimulus activity,
as in Eq.~(\ref{m0v}), implying that there are approximately $70\%$
receptor dimers are at the state corresponding the lower rate
of CheA autophosphorylation. 
Although the data on  response-stimulus relation 
are  very limited,   they are used to  estimate that 
$\beta B/(1-\beta\nu J)\approx 10.49$,  
which compares the effect of ligand binding with that   of cooperation. 
We study adaptation time for two different sets of 
data~\cite{spudich,jasuja1999a}. Assuming the delayed time in feedback 
to be  $0.1s$,  it is   found that 
the feedback strength compared with coupling, $\beta\sigma/(1-\beta\nu J)$,
is approximately $0.0068$ to $0.013$, or $0.0015$ to $0.0047$, respectively.
These numbers  obtained from different data and  by  using different methods
are of the same order of magnitude,
as  a   sign of the consistency of theory. 
Precise information on the feedback delay 
 time can improve this determination. 
From the data on the relative CheA activity~\cite{bornhorst},
 we obtain Eq.~(\ref{fe}), which gives the  relation
between the two levels of CheA activity corresponding to the two
conformations of the  receptor dimer.
Combined with other results,  it tells that  
the ratio between the two levels of CheA activity is
$A_{-1}/A_{1}\approx 164.77$,   in good  consistency with the 
available experimental information on this  ratio. We note that 
the fitting  is not perfect. This may be  partly due to 
the simple nature of the minimum model and
further simplified treatment, and partly due to insufficient experimental
information.  However, with a working framework proposed, we anticipate more
experimental and theoretical discoveries stimulated by  the current 
attempt. On the other hand,   it would not be satisfactory  to us to
have  a good fitting of  the data by simply tuning parameters 
without a  clear physical picture.   

We  need  improvement on the  available  
experimental results,  as well as new  experimental information, 
to provide a basis for    extension  and refinement of   the 
theory. 
For example,  we need a significant broadening  of the range of 
occupancy change in response-stimulus relation. We also need a
clearer relation between  adaptation time and occupancy change. 
The relation between CheA activity and the receptor conformational state 
and CheA activity, as well as the relative rate of the two rotation modes, 
is vital for  going  beyond the simple treatment here. 
More accurate results on  $A_{-1}/A_1$ is also important.
Independent determination of the dissociate constant is of fundamental
importance.  
Most exciting experiments might be  direct measurements of the conformational
states $V^0$, $V^1$, and the coupling coefficient $T_{ij}$, as well
as the  energy change or force induced by the ligand binding.
A clarification on whether the conformational change is rotation or
a vertical displacement is interesting. For the former,
 $V^0$ and $V^1$ are angles,
while $H$, the effect of ligand binding, is a torque.
For the latter,   $V^0$ and $V^1$ are positions, while  $H$ is a force.
The receptor-receptor interaction can be determined by 
measuring the relation of force or torque on one receptor dimer and
the conformations of its neighbors. 
This would be  a direct test of the
conformation-dependent interaction. A determination of the geometry of the 
lattice is also interesting, from which one  can obtain
the value of  $\beta\nu J$, and consequently other 
parameter values.  

Our theory is  entirely different from Hill model. An integer Hill
coefficient is understood as the number of ligands bound to a receptor.
A  non-integer Hill coefficient, as often used, 
 does not seem clear conceptually though could be tuned to fit the data. 
Nonetheless,  
from mean field point of view, the effect of receptor-receptor
interaction  could be viewed as effective additional ligand binding.
Therefore from this perspective, 
the conclusion on limited cooperativity in \cite{bornhorst}
is consistent with strong thermal fluctuation in our theory. 

Here we specialize in chemotactic receptors, however, the theory may also 
apply to many other receptor systems.  For example, 
state-dependent co-inhibition 
between transmitter-gated cation channels was observed~\cite{khakh}.
Clustering of $GABA_{A}$ receptors and
the decrease of affinity was also studied~\cite{chen}, which was also 
 analyzed in terms of Hill model in a similar way to \cite{bornhorst},
 thus it  can also 
be explained by our theory as  an indication of 
receptor-receptor interaction and  thermal fluctuation. 
In many receptor systems, clustering, or called oligomerization, 
together with signalling, occurs as a response to stimulus. 
Theoretical investigation on this situation is  presented  elsewhere.

In finishing this paper,
let us  list some new   experiments anticipated from the 
point of view of this theory. 
(1) More clarifications on conformational change induced by ligand
binding, and determination of conformational change due to interaction with
another receptor dimer.
(2) Direct measurement of the forces generated by ligand binding and
 by conformational change of the neighboring receptor dimer. 
(3) Independent determination of dissociate constant using other methods.
(4) Investigations on the responses corresponding to fixed 
conformational states, and hence  $r_1$ and $r_{-1}$ as discussed above.
(5) Direct measurements on CheA and CheY activities.
(6) More clarification on the relation between the receptor conformational
 state and CheA activity.
(7) Increasing the range of occupancy change in response-stimulus relations,
and more accurate determination  of pre-stimulus occupancy and occupancy
change.
(8) More accurate determination of  adaptation time as a function of 
 the occupancy change. 
(9) Precise determination of the form of  energy function. 
(10) Determination of the details of feedback due to
the change of the methylation level, including the delay time.

\newpage

Figure captions:

Fig.~1. An illustrative snapshot of the configuration of
receptor dimers on a $50\times 50$ square lattice. 
An  up triangle  represents 
the conformation state $S_i=1$, a down triangle represents $S_i=-1$,
a filled triangle represents  ligand binding, an empty triangle represents
no ligand binding. 

Fig.~2. Response-stimulus relation $\delta m$ vs. $\delta c$.
The data points are transformed from those read from
\cite{jasuja1999b} with a computer software. The range of
receptor occupancy change is too small, so only qualitative comparison 
is possible.
The straight line is the least square fitting
$\delta m=10.49 \delta c$.

Fig.~3. (a) Normal-normal plot of the 
relation between adaptation time $t^*$ and occupancy change
$\delta c$.  The data points are adopted from \cite{spudich} 
by using a computer software, 
 with  the
concentration  transformed to occupancy.
(b) Normal-log plot of the same data, showing that they can be fitted to 
a logarithmic relation. 

Fig.~4. Relation between adaptation time $t^*$ and occupancy change
$\delta c$. 
The data points are adopted from \cite{jasuja1999a}
by using a computer software, 
 with  the
concentration  transformed to occupancy. The straight line is the 
 least square fitting
$t^*=156.3513\log_{10}\delta c+114.9912$.

Fig.~5. The relation between the relative CheA autophosphorylation rate
$R$
and ligand concentration $L$. The data points are adopted from 
\cite{bornhorst} by using a computer software.
 The theoretical  curve  is $R=1-F\frac{L}{L+K_d}$, with
$F=0.95$ and $K_d=20\mu M$. 

\psfig{figure=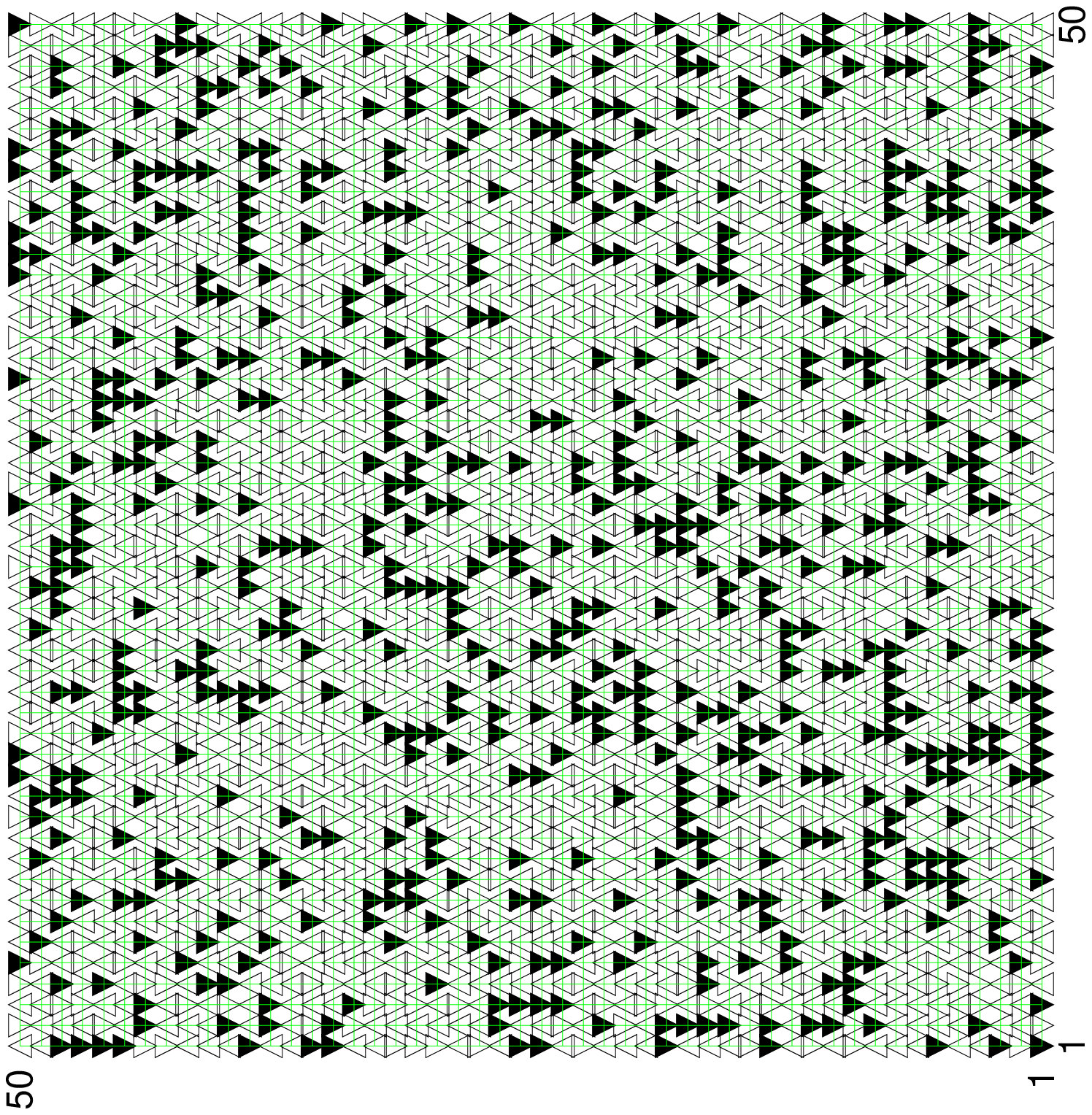}
\psfig{figure=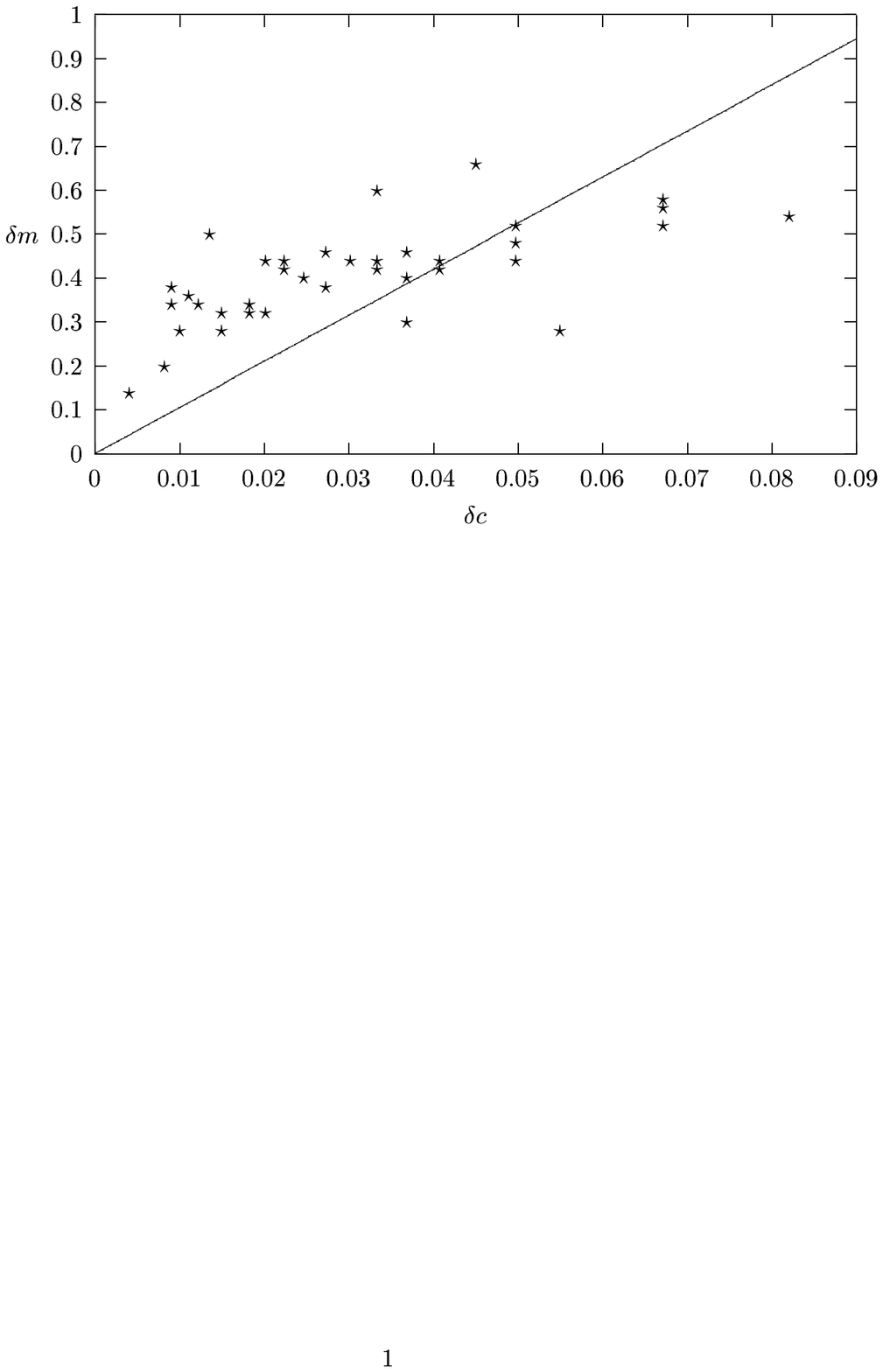}
\psfig{figure=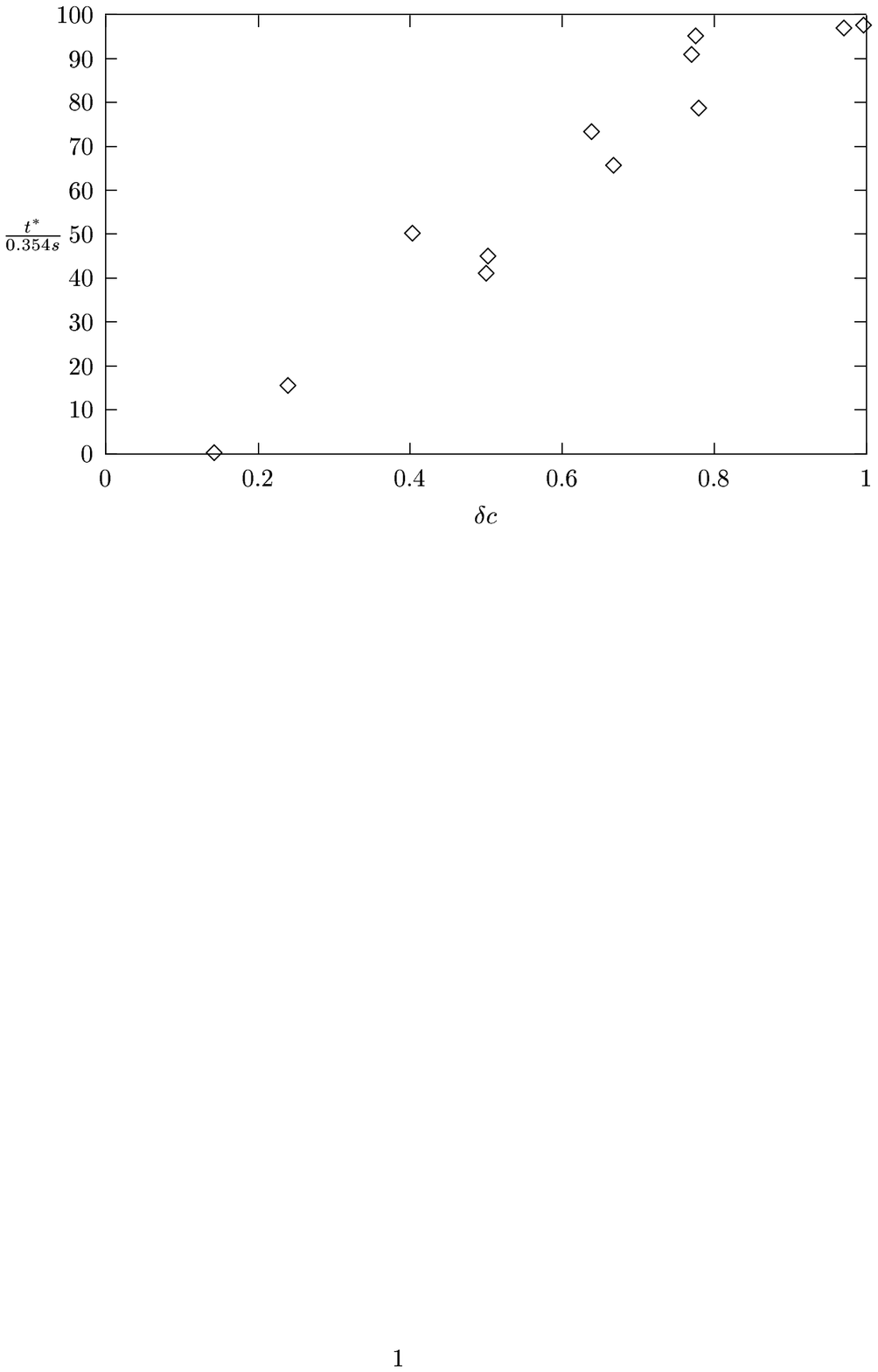}
\psfig{figure=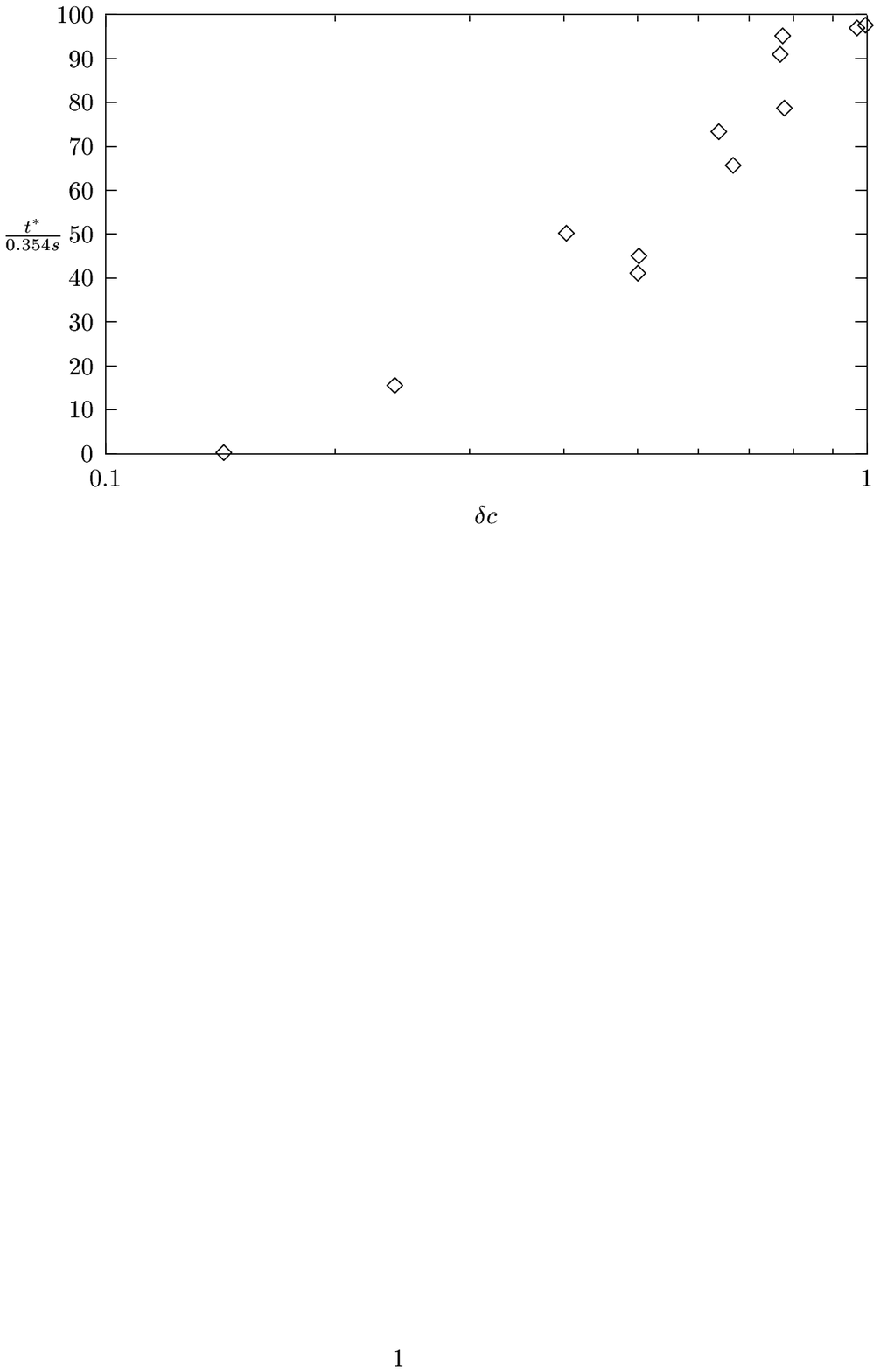}
\psfig{figure=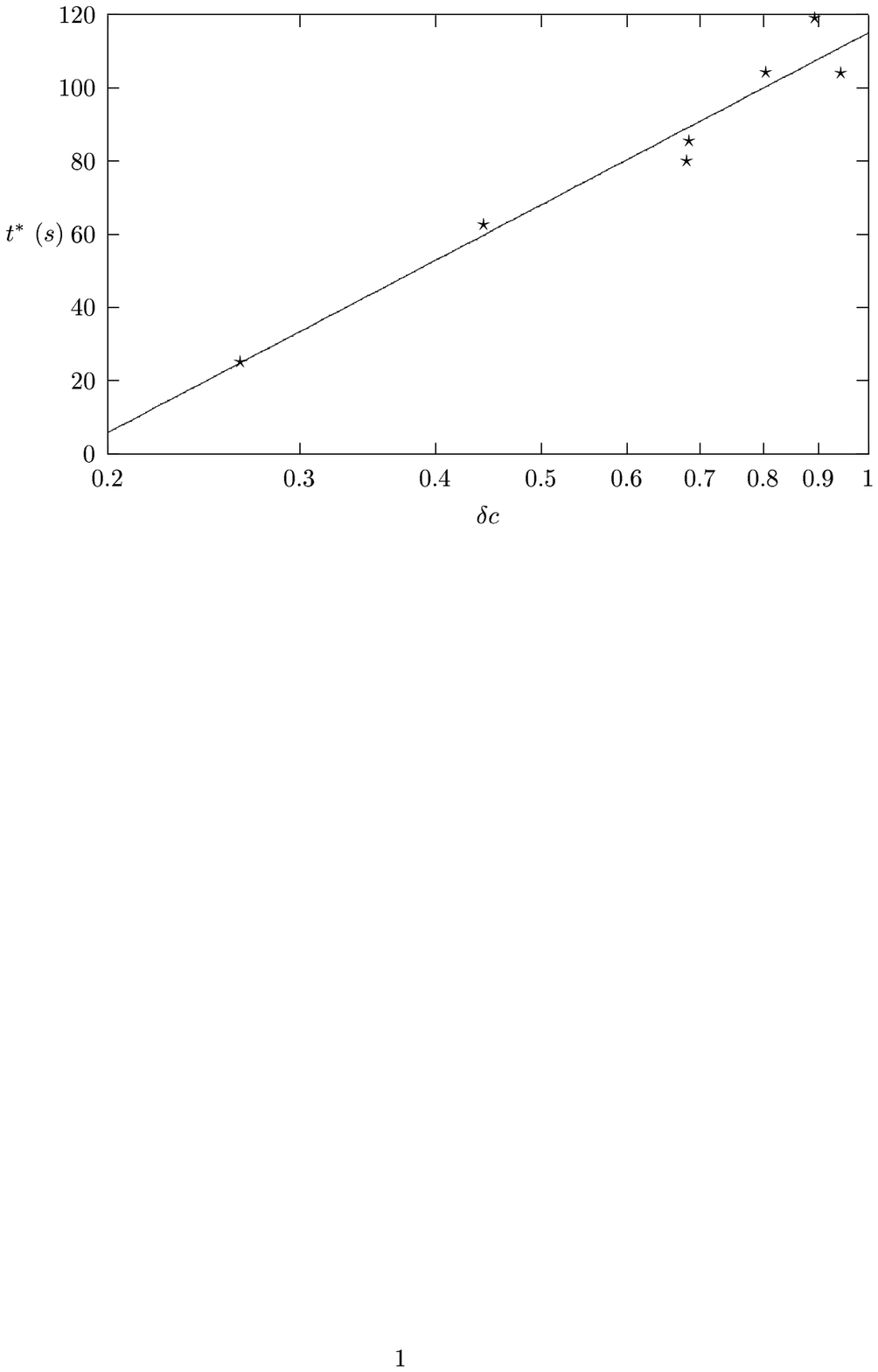}
\psfig{figure=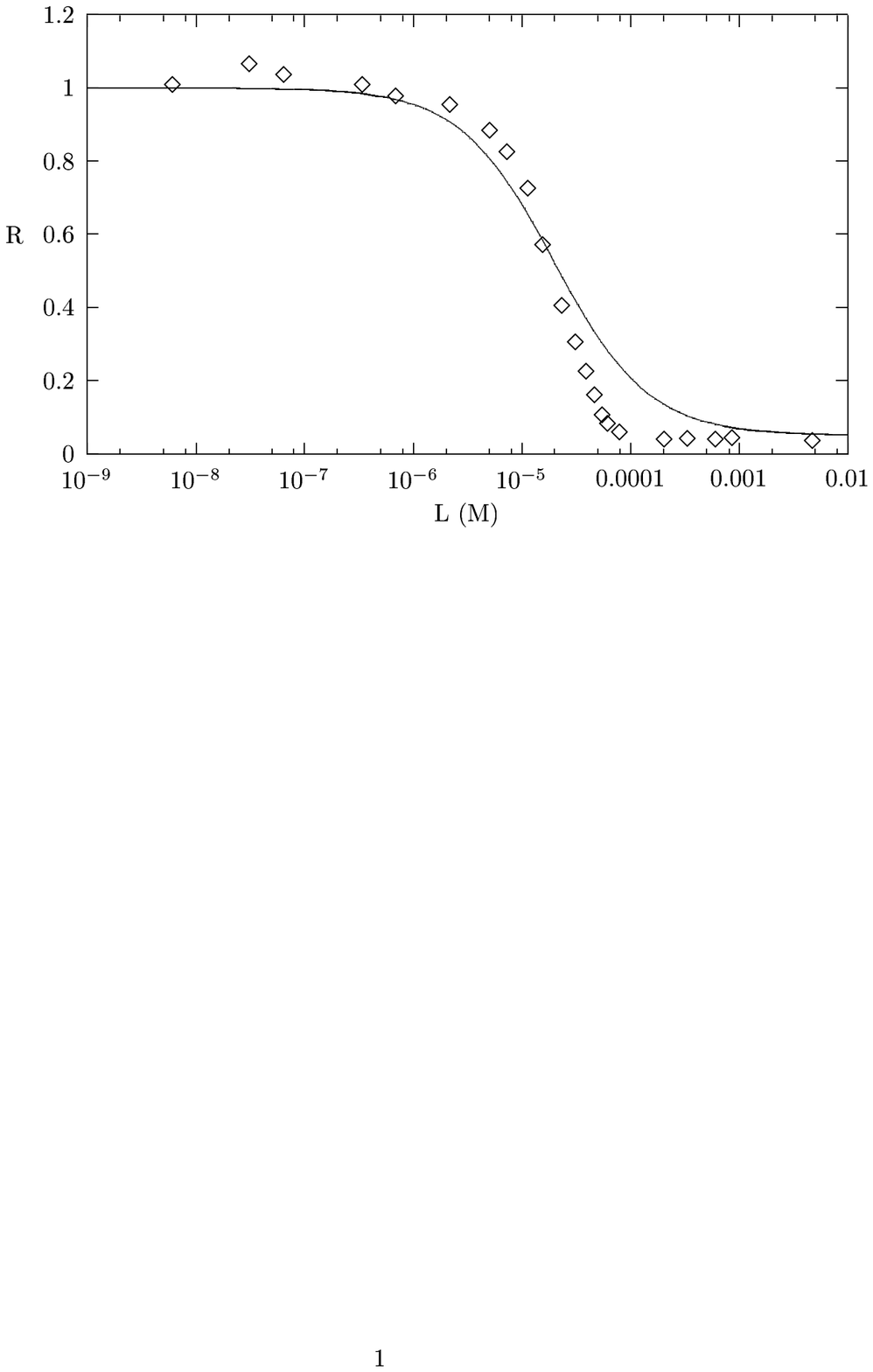}


\begin{references}

\bibitem{berg1993}
H. Berg, 
{\em Random Walks in Biology}
 (Princeton University Press, Princeton, 1993).

\bibitem{stock} J.  Stock and  M. Surette, 
 {\em Escherichia coli and
 Salmonella typhimurium: Cellular and Molecular Biology}, ed. F.C. Neidhardt,
(ASM, Washington, 1996).

\bibitem{falke}
J. J. Falke {\it et al.}, Ann. Rev. Cell Dev. Bio. {\bf 13}, 457 (1997).

\bibitem{berg}
H. Berg,    Phys. Today 
 {\bf 53}(1), 24  (2000).

\bibitem{gardina} I. Tatsuno, M. Homma, K. Oosawa, and I. Kawagishi,
  Science {\bf  274}, 423 (1996); P. J. Gardina and M. Manson, 
Science {\bf 274}, 425 (1996).  

\bibitem{segall}  S. M. Block, J E. Segall and H. C. Berg,
 J. Bacteriol. {\bf 154}, 312 (1983).
J E. Segall, S. M. Block and H. C. Berg, Proc. Natl. Acad. Sci. USA {\bf 83}, 
8987 (1986).

\bibitem{borkovich} K. A. Borkovich, L. A.  Alex and M. Simon, 
 Proc. Natl. Acad. Sci. USA  
 {\bf 89}, 6756 (1992).

\bibitem{parkinson} J.S. Parkinson and D.F. Blair,
  Science  {\bf 259}, 1701 (1993); M.R.K. Alley, 
J.R. Maddock,  and L. Shapiro, Science  
 {\bf 259}, 1754 (1993); 
J.R. Maddock and L.  Shapiro,  Science   {\bf 259}, 1717  (1993).

\bibitem{bray} D. Bray, M. D. Levin and C. J.  Morton-Firth, 
 Nature  {\bf 393}, 85 (1998). 
In the so-called raindrops calculation therein, it was not considered 
that  each raindrop was actually an  aggregate,  rather than a set
of random sites. This calculation,   if done
in a right way, would  not give the ``maximum concentration''
as expected there.
It would only  be  an improvement on the ``minimum'' one in Eq. (1) therein,  
which was actually obtained without 
considering  possible  overlap  between  ``raindrops''.
Since most of the chemotactic receptors form a
single cluster,  the idea  there would  
lead to very close or a same magnitude of 
response for  different values of occupancy. 

\bibitem{berg2}
H. Berg and L. Turner, 
 Proc. Natl. Acad. Sci. USA  {\bf 92}, 447 (1975).

\bibitem{shi1} Y. Shi   and T.  Duke,
 Phys. Rev. E  {\bf 58}, 6399 (1998);  also obtainable at
http://xxx.arXive.org/abs/physics/9901052. We did not 
draw   a picture of a lattice there, but  it is apparent
that   a lattice model  is discussed. 

\bibitem{shi2} Y. Shi, 
  Europhys. Lett. {\bf  50}, 113 (2000);  also obtainable at 
http://xxx.arXive.org/abs/physics/9901053.  

\bibitem{jasuja1999b} R. Jasuja, Y.  Lin, D. R. Trenthan and S. Khan, 
Proc. Natl. Acad. Sci. USA
 {\bf 96}, 11346 (1999).

\bibitem{bornhorst}
J. A. Bornhorst  and J. J. Falke, Biochemistry 
  {\bf 39}, 9486 (2000).

\bibitem{sourjik} 
V. Sourjik and H. C. Berg, 
Mol. Microbio. {\bf 37}, 740 (2000).

\bibitem{hill} A. V. Hill, 
J. Physiol. {\bf 40}, iv (1910).

\bibitem{ying} 
C. C. Ying  and F. A. Lai,
Nature Cell Bio.  {\bf 2}, 669 (2000).

\bibitem{bieman}
H. P. Bieman  and D. E. Koshland, 
 Biochemistry  {\bf 33}, 629 (1994).

\bibitem{alon} U. Alon, M. G.  Surette, N. Barkai and S. Leibler
  {\bf 397}, 168 (1999).

\bibitem{jasuja1999a} R. Jasuja  {\it et. al},
 Biophys. J.  {\bf 76}, 1706 (1999).

\bibitem{berg3}
H. Berg  and P. M. Tedesco,  Proc. Natl. Acad. Sci. USA  {\bf 72}, 
3235 (1975).

\bibitem{spudich} J. L. Spudich  and D. E. Koshland, 
 Proc. Natl. Acad. Sci. USA,  {\bf 72}, 710 (1975).

\bibitem{khan1993}
S. Khan  {\it et. al},  Biophys. J.  {\bf 65}, 2368 (1993).

\bibitem{khakh}   B. S. Khakh, {\it et. al} 
 Nature, {\bf 406}, 405 (2000).

\bibitem{chen}
L. Chen, H. Wang, S. Vicini and R. W. Olsen,  
Proc. Natl. Acad. Sci. USA,  
 {\bf 97},  11557 (2000).


\end{references}
\end{document}